\documentclass[epj]{svjour}
\usepackage{graphics}
\begin{document}
\title{Charge orderings in the atomic limit of the extended Hubbard model}
\author{G. Paw{\l}owski}
\institute{Institute of Physics, A. Mickiewicz University, ul. Umultowska 85, 61-614 Pozna\'{n}, Poland}
\date{}
\abstract{
The extended Hubbard model in the atomic limit (AL-EHM) on a square lattice with periodic boundary conditions is studied with use of the Monte Carlo (MC) method. Within the grand canonical ensemble the phase and order-order boundaries for charge orderings are obtained. The phase diagrams include three types of charge ordered phases and the nonordered phase. The system exhibits very rich structure and shows unusual multicritical behavior. 
In the limiting case of ${t_{ij}=0}$, the EHM is equivalent to the pseudospin model with single-ion anisotropy $\frac{1}{2}U$, exchange interaction $W$ in an effective magnetic field ${\left( \mu - \frac{1}{2}U - z_{0}W\right) }$. This classical spin model is analyzed using the MC method for the canonical ensemble. The phase diagram is compared with the known results for the Blume-Capel model.
} 
\maketitle
\section{Introduction}
After discovery of high temperature superconductors, evidencing the existence of the charge density waves (CDW) in such materials, the research work on charge orderings has been in the main flow of condensed matter physics. The problem of reciprocal competition and coexistence of CDW phase and other orderings has been widely studied in literature (\cite{Gabovich} and references therein). The charge ordering of strongly correlated electron systems has recently attracted much attention also, due to a considerable effect on the colossal magnetoresistance observed in the manganese perovskite compounds ($La_{1-x} Sr_x MnO_3$, $Pr_{1-x} Ca_x MnO_3$) \cite{Goto}. It should be emphasized that charge order (CO) was observed not only at commensurable lattice filling, but also in a broad doping regime \cite{Reis}.

The adequate model for description of charge-ordered electron systems is the extended Hubbard model (EHM) \cite{Rob_73}-\cite{Zang}. Detailed analysis of the model has been made for a special case of quarter- or half-filled band, although there are known results for the whole range of electron concentration \cite{Hoang}-\cite{Onari}, but this general case has been relatively poorly examined. The basic limiting version of EHM, suitable for description of CO, is the case of the very narrow-band, when $t_{ij}\ll U,W$. We can go to the classical limit of the model and put $t_{ij} = 0$ (so-called extended Hubbard model in the atomic limit or UV model). Many authors have studied critical behavior of this model \cite{Bari}-\cite{Mancini} and the staggered charge order was observed irrespectively of the method used.
On the other hand, the experimental observations of the CO insulating phase in many compounds with very narrow-band ($BaBi_x Pb_{1-x} 0_3$, $Cs_2 SbCl_6$, $Fe_3 O_4$, $CsAuCl_3$ and several complex TCNQ salts; \cite{Micnas} and references therein) motivate to theoretical research on the extended Hubbard model in the atomic limit.

We present here the results of MC calculations for 2D AL-EHM on a square lattice with periodic boundary conditions (PBC). We obtain the phase diagrams in the whole range of electron concentrations $\{0, 2\}$ and determine the order-order boundaries.

The Hamiltonian of AL-EHM has the form
\begin{eqnarray}
{\cal H}=U{\sum_{i}n_{i\uparrow} n_{i\downarrow}}+W\sum_{ij}n_i n_j- \mu
\sum_i n_i ,
\label{UW}
\end{eqnarray}
where $n_i = n_{i\uparrow} + n_{i\downarrow}$ is the number of electrons on the $i$th site $\{0, \uparrow, \downarrow, \uparrow \downarrow\}$, $U$ represents the on-site and $W$ intersite Coulomb interaction (in this paper we assume the $U, W \geq 0$ case only) and $\mu$ is the chemical potential.

The structure of this paper is as follows. In Section 2 we discuss details of the simulation methods we use, the statistical ensembles and give the criteria for critical and order-order transitions. Using the grand canonical ensemble (GCE) we present the way to obtain $n(\mu)$ curve ($n$ is the total electron concentration), which is crucial for the study.
Much attention is devoted to the 'contour cluster' defined for ordered phases. We present a percolation probability analysis for defined clusters.

In Section 3 we present results for AL-EHM eq. (\ref{UW}). Evidence is provided for the existence of three types of CO phases and the nonordered (NO) phase. The system exhibits the first- and second-order transitions and tricritical points. The phase diagram also presents the order-order boundaries.

In Section 4 the equivalence between the model (\ref{UW}) and the pseudospin model with  double 'zero' state degeneracy in the magnetic field is proved. We present also the transformation to the well-known Blume-Capel (BC) model with a temperature-dependent single-ion anisotropy. Our results are in very good compatibility to a detailed analysis of BC model performed by the MC method \cite{Kimel}-\cite{Xavier}. The presentation of order-order structural boundaries is new in our analysis (analogous calculations for BC model have been given in our recent paper \cite{gpawlo} using the Wang-Landau flat histogram algorithm).

In the Appendix we present a comparison of our results with those obtained by analytical methods, mean field (MFA) and Bethe-Peierls (BPA) approximation.

\section{Details of the simulation methods}

To simulate the behavior of AL-EHM we use the grand canonical ensemble. In this distribution the model analyzed describes for example adsorption of electron gas with the maximum double site coverage on a lattice. In a general case the model is applicable to a two-component lattice gas \cite{Halvorsen}.

A long-time GCE Monte Carlo (GCMC) simulation \cite{Landau} is adopted for relatively small systems (L=10-40) described by model (\ref{UW}), while for large systems (L=60-80) we analyze the pseudospin model (cf. eq. (\ref{eq:3})) using the canonical Monte Carlo. The results of both methods are complementary.

In our simulation we use a local update method determined by elementary \textit{runs} \cite{Landau}. GCMC algorithm includes three sequential update steps:\\
\textit{insertion of the particle}: we randomly choose the spin direction (up/down); we randomly choose a site; if the local register of spins in this direction is empty we accept the motion with a proper probability (depending on the number of particles in the system, the interactions and temperature); and we introduce the particle into the register;\\
\textit{removal of the particle}: we randomly choose a particle from the register; we accept its removal from the register with a proper probability; we clear out the register cell and the last position of the register we move to the emptied place to ensure the register continuity;\\
\textit{move of the particle} (the canonical part): we randomly choose a particle from the register, randomly choose a site and check if there is an electron with the same spin at the site, if it is not there we accept the move with a standard Metropolis probability, and we change the register record. In general, this procedure can be omitted being a simple combination of \textit{removal} and \textit{insertion}.

\begin{figure}
\includegraphics{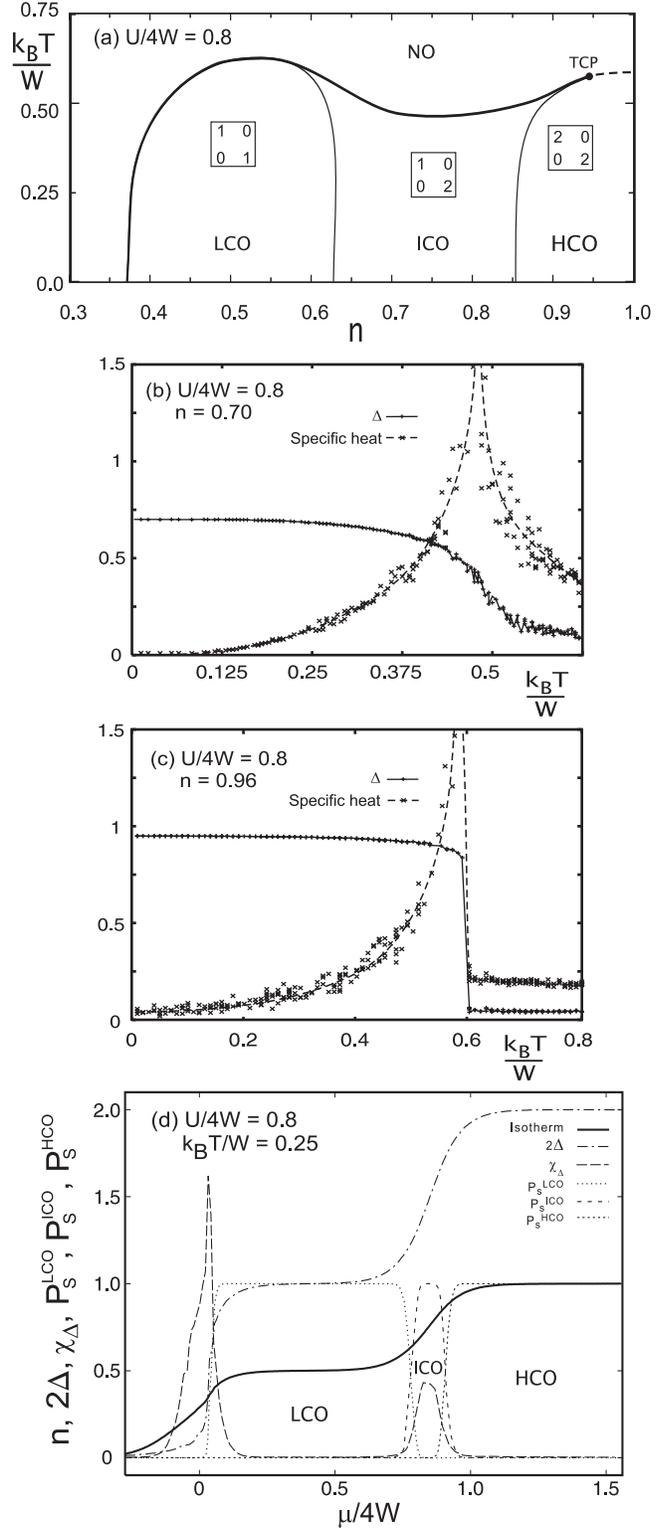}
\caption{Example of simulation results. (a) phase diagram for $U/4W = 0.8$, the second- and first-order transitions are indicated by the solid and the dashed curves, respectively. The thin lines indicate order-order boundaries. Detailed analysis performed for selected parameters is presented below. (b)(c) charge order parameter $\Delta$ and specific heat as a function of temperature for fixed concentrations ($n=0.70;0.96$). (d) adsorption isotherm $n(\mu)$, charge order parameter $\Delta$, CO susceptibility $\chi_\Delta$ and percolation probabilities $P^{(phase)}_S$ of the ordered structures for $k_B T/W = 0.25$. Other denotations in the text.}
\label{U4W08}
\end{figure}
 
For the spin model we use a standard Metropolis update (the simulations for the canonical case we have used for comparative purposes). More recent MC methods for spin models (e.g. BC model) are presented by Loison \textit{et al.} \cite{Loison}.

To analyze the charge ordered states, we take into account the on-site and nearest-neighbor intersite interactions. We divide the square lattice into two interpenetrating sublattices A, B and define the charge order parameter as $\Delta = \frac{1}{2} | n_A - n_B |$, where $n_A, n_B$ are the electron concentrations on sublattices A and B, respectively.

Detailed analysis of the thermodynamic characteristics (such as the order parameter, internal energy, specific heat, charge order susceptibility) has been performed (i) as a function of the chemical potential $\mu$ for a fixed temperature or (ii) as a function of temperature $k_BT$ for a fixed electron concentration (Fig. \ref{U4W08}). Such an analysis brings about the range of existence of the CO phase. As our analysis is limited to a restricted number of sites of the system we observe the finite-size effects on the order parameter (the finite value of $\Delta$ over critical temperature, cf. Figs \ref{U4W08}b,c). Precise location of the critical points has been determined by the discontinuity of the specific heat and the charge order susceptibility. 

The most essential characteristic of our analysis is the adsorption isotherm defined as the average site occupation as a function of the chemical potential $n(\mu)$, which allows as to plot phase diagrams as the function of electron concentration. For each value of $\mu$ we make about $\sim 10^4$ Monte Carlo steps (MCS) for the 2nd-order transition and about $\sim 10^5 - 10 ^6$ for the 1st-order one.

In the case of 1st-order transition the system described by GCE reaches equilibrium after a very large number of simulation steps. Analysis of a discontinuous transition is difficult, therefore, we adopt the histogram method also.

It is interesting to analyze structural details of charge ordered states. In our simulation we analyze the existing clusters for each generated sample, using the cluster finding routine \cite{Hoshen}. It turns out that in all ordered states we can find a spanning cluster with some staggered structure.
In AL-EHM there are three different, elementary and energetically stable charge ordered structures (cf. Fig.  \ref{GS_snapshot}):\\
LCO - Low Charge Order ($1010$ or $1212$),\\
HCO - High Charge Order ($2020$),\\
ICO - Intermediate Charge Order ($x0y0$ or $x2y2$, where $x,y = \{1,2\}$, excluding the case when for whole ordered domain $x = y$).\\
The first - Low Charge Order is realized for a checkerboard combination of electrons (holes) and empty (full) sites; the second - High Charge Order refers to electron pairs and empty sites. The last charge ordered state is a simple combination of LCO and HCO structures, therefore this state can be called the Intermediate Charge Order.
Beyond these three orders we observe the disordered states denoted as NO.

\begin{figure}
\includegraphics{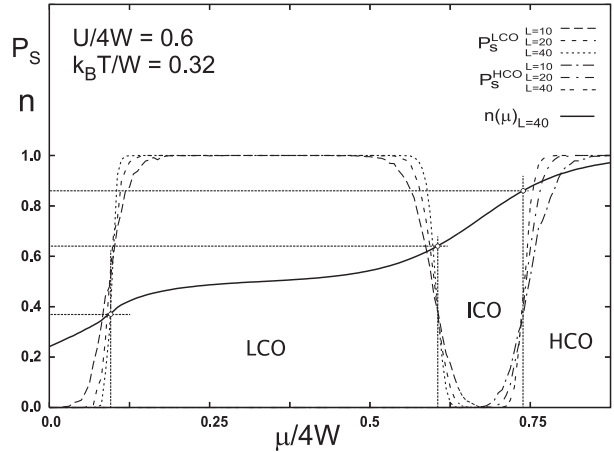}
\caption{The percolation probabilities of the ordered phases for different lattice sizes (the exemplary plot for $U/4W = 0.6$ and $k_BT/W = 0.32$). The percolation thresholds for infinite lattice for LCO $\leftrightarrow$ ICO $\leftrightarrow$ HCO sequence are obtained by the intersection method.}
\label{ph-th}
\end{figure}

The exact ground state analysis for AL-EHM \cite{Borgs} shows that there are two different charge ordered states (in our notation: LCO and HCO). Similar results for the Blume-Capel model \cite{Kimel} give two AF phases (1,-1) and (0,1) in the ground state. Until now it has been an open question which kind of ordered structures exist at finite temperatures in the models considered. No exact evidence for the existence of finite-temperature order-order boundaries in AL-EHM \cite{Jedrzejewski} as well in the equivalent Blume-Capel model \cite{Kimel} has been found yet. The mean-field results show that such internal boundaries are present in a limited range in finite-temperatures \cite{Micnas}. It is very interesting to analyze this problem again. For $T > 0$ we can show the existence of ordered structures (especially the \textit{infinite} contour clusters) which are present in the ground state as pure phases.
In our work we have made the percolation probability analysis. Such a quantity ${P_S}^{(phase)}$ is defined as a fraction of conducting samples (in the sense of a \textit{spanning cluster} of \textit{phase} occurring in the sample) during one cycle of simulation. The percolation transition points are obtained as abscissa of the points of intersection of a family of the percolation curves for different finite-size results (the intersection method \cite{Kirkpatrick}). Thus defined percolation threshold is very well determined in the whole range of the model parameters (Fig. \ref{ph-th}).

\section{Review of the results}
\label{3}
Here we present results in the whole range of electron concentrations $0 \leq n \leq 2$. Since Hamiltonian (\ref{UW}) exhibits the electron-hole symmetry some diagrams are presented in the $|n - 1|$ symmetrical plane.

\subsection{Ground state phase diagram}
\label{3.1}

\begin{figure}
\includegraphics{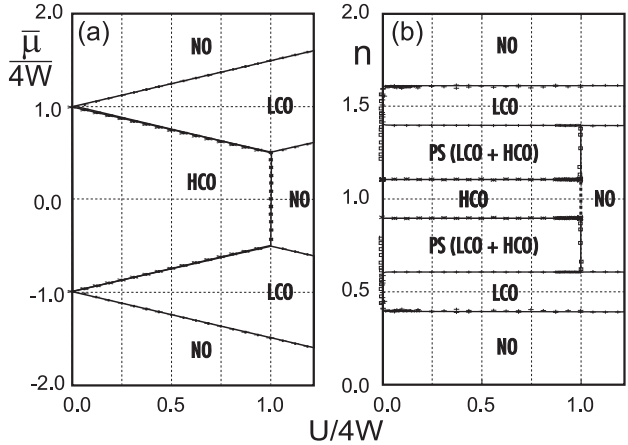}
\caption{The ground state phase diagrams (extrapolated from $k_B T/W = 0.01$) as a function (a) of the reduced chemical potential $\bar{\mu}/4W=(\mu-\frac{1}{2}U-4W)/4W$ and $U/4W$, (b) of the electron concentration $n$ and $U/4W$. Denotations: NO - nonordered phase, PS - phase separation of LCO and HCO.}
\label{GS_pd} 
\end{figure}

\begin{figure}
\includegraphics{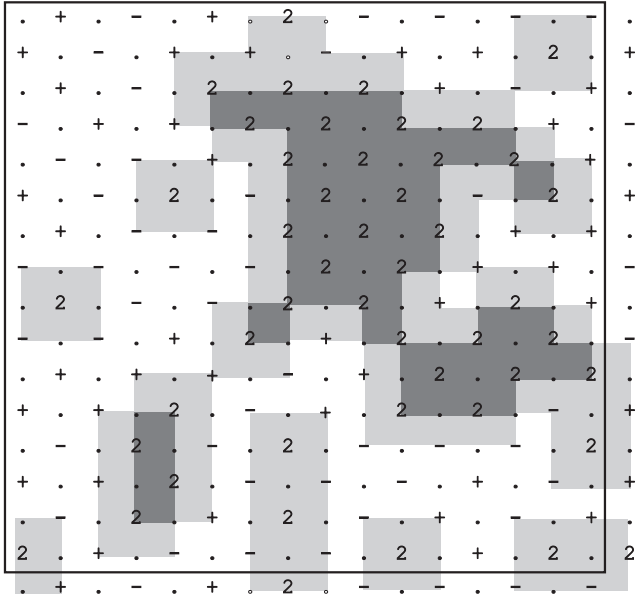}
\caption{The Monte Carlo snapshot of the ordered state near the ground state ($k_B T/W = 0.0125$, $U/4W = 0.6$, $n = 0.68$) for square lattice $L = 16$ (with respect to the periodic boundary condition). The overall checkerboard charge order include the 'pure' LCO state (white area) and the separated HCO domains (the dark gray area). Phase separation of LCO and HCO states is observed in the range of the parameters corresponding to the jumpwise change in the electron concentration as a function of chemical potential.}
\label{GS_snapshot}
\end{figure}

The left part of Fig. \ref{GS_pd} presents the phase diagrams of model (\ref{UW}) as a function of the reduced chemical potential $\bar{\mu}/4W=(\mu-\frac{1}{2}U-4W)/4W$ and the model interaction ratio $U/4W$ for the extrapolated ground state (GS). We observe only two staggered charge ordered phases (LCO, HCO) and the NO phase. The linear character of phase boundaries between phases is identical with that concluded from the analytical results \cite{Micnas} - \cite{Jedrzejewski}.
The phase transitions $LCO \leftrightarrow NO$ is 2nd-order, while the $HCO \leftrightarrow NO$ and the $LCO \leftrightarrow HCO$ is discontinuous (1st-order). For $\bar{\mu}=0$ the $HCO \leftrightarrow NO$ is the transition from the CO staggered state (2020) to the Mott state (1111).

The right part of Fig. \ref{GS_pd} presents the corresponding phase diagrams as a function of electron concentration. We observe simple linear boundaries between phases also, but the LCO-HCO transition does not take place here. For some range of electron concentrations there is an area between the LCO and HCO states with a constant value of chemical potential (for fixed $U/4W$) which leads to a state with phase separation (PS). The state with phase separation of CO has been observed also for EHM in the case $t \neq 0$ \cite{Tong}.

The range of electron concentrations with a stable LCO phase $(0.37,0.63)$ and $(1.37,1.63)$ remains the same irrespectively of the value of $U/4W$ (the result is obvious due to the fact that the (1010) order does not include electron pairs). The critical points for $U\rightarrow \infty$ of LCO phase are in a good agreement with the percolation limit in the lattice gas (see Appendix).

Much more unexpected is the invariability of HCO boundaries as a function of on-site interaction. The concentration range of HCO $(0.86,1.14)$ remains the same for $0 \leq U/4W \leq 1$.

\subsection{\label{3.2}Finite temperatures}

\begin{center}
i) Half-filling case.
\end{center}

\begin{figure}
\includegraphics{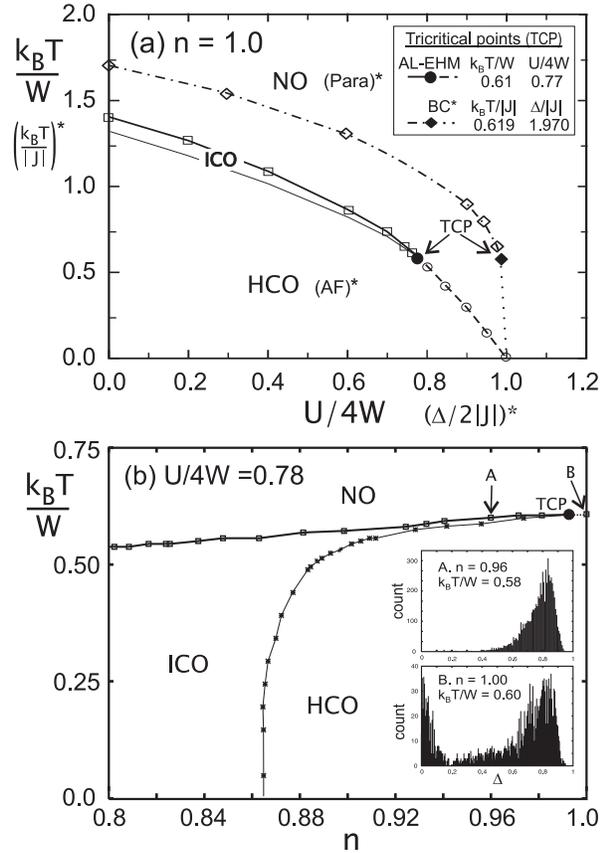}
\caption{(a) The phase diagram for AL-EHM at $n=1$ and BC model at $H=0$ (MC, *after \cite{Kimel}). Comparison of the location of the tricritical point (TCP). (b) Phase diagram near $ n = 1$ for $U/4W = 0.78$ (GCMC). Small boxes present histograms of order parameter for (A) second order ($n = 0.96$, $k_B T/W = 0.58$) and (B) first order phase transition ($n = 1.00$, $k_B T/W = 0.60$). }
\label{n_1} 
\end{figure}

The case of $n = 1$ is well known {\cite{Bari}-\cite{Halvorsen}}. It is a particular case for which we know the analytical relation between the chemical potential and the interaction terms $\mu=\frac{1}{2}U+z_0 W$ for any dimension ($z_0$ is the number of nearest-neighbors). Consequently, for the half-filling case we can use the canonical simulation. Fig. \ref{n_1}a presents the results for the AL-EHM model for $n = 1$ and the BC model in a suitable limit when the external magnetic field is zero (the rhomb represents the critical points between AF and paramagnetic order). A detailed comparison of the results for both models is made in Sec. IV.

At the half-filling we observe two types of ordered phases (HCO, ICO) and NO state. For $U/4W > 1$ the system is in the disordered Mott state. In the range $0.77 < U/4W \leq 1$ there is a discontinuous (1st-order) transition between CO (HCO) and NO phase. At the point $U/4W = 0.77$, $k_B T/W = 0.61$ we observe tricritical (TCP) behavior of the system. At this point the line of the phase transition and the order-order boundary join. For $U/4W < 0.77$ the transition becomes a 2nd-order (we observe the sequence HCO $\leftrightarrow$ ICO $\leftrightarrow$ NO).

\begin{center}
ii) The case of $0 \leq n \leq 2$.
\end{center}

\begin{figure}
\includegraphics{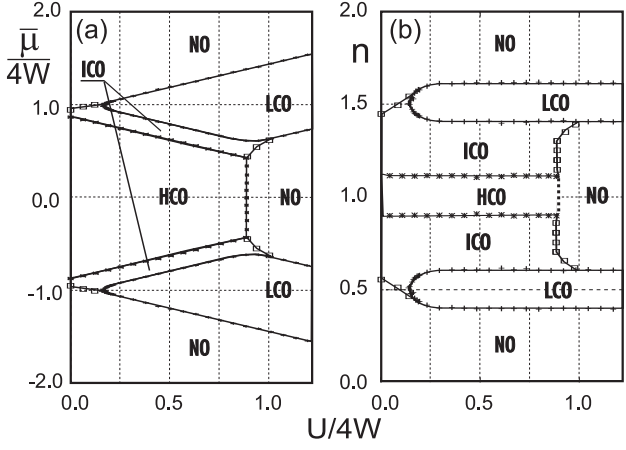}
\caption{Phase diagrams as a function (a) of reduced chemical potential $\bar{\mu}/4W=(\mu-\frac{1}{2}U-4W)/4W$ and $U/4W$, (b) of the electron concentration $n$ and $U/4W$, plotted for $k_B T/W = 0.25$. Denotations as in Fig. \ref{GS_pd}.}
\label{t025_pd} 
\end{figure}

In low temperatures the phase diagram remains qualitatively the same as in GS (Fig. \ref{t025_pd}). The PS state changes into the ICO state (which is in fact a mixed phase of the LCO and the HCO states). The phase boundaries loose their linear character but the type of order remains the same.

\begin{figure}
\includegraphics{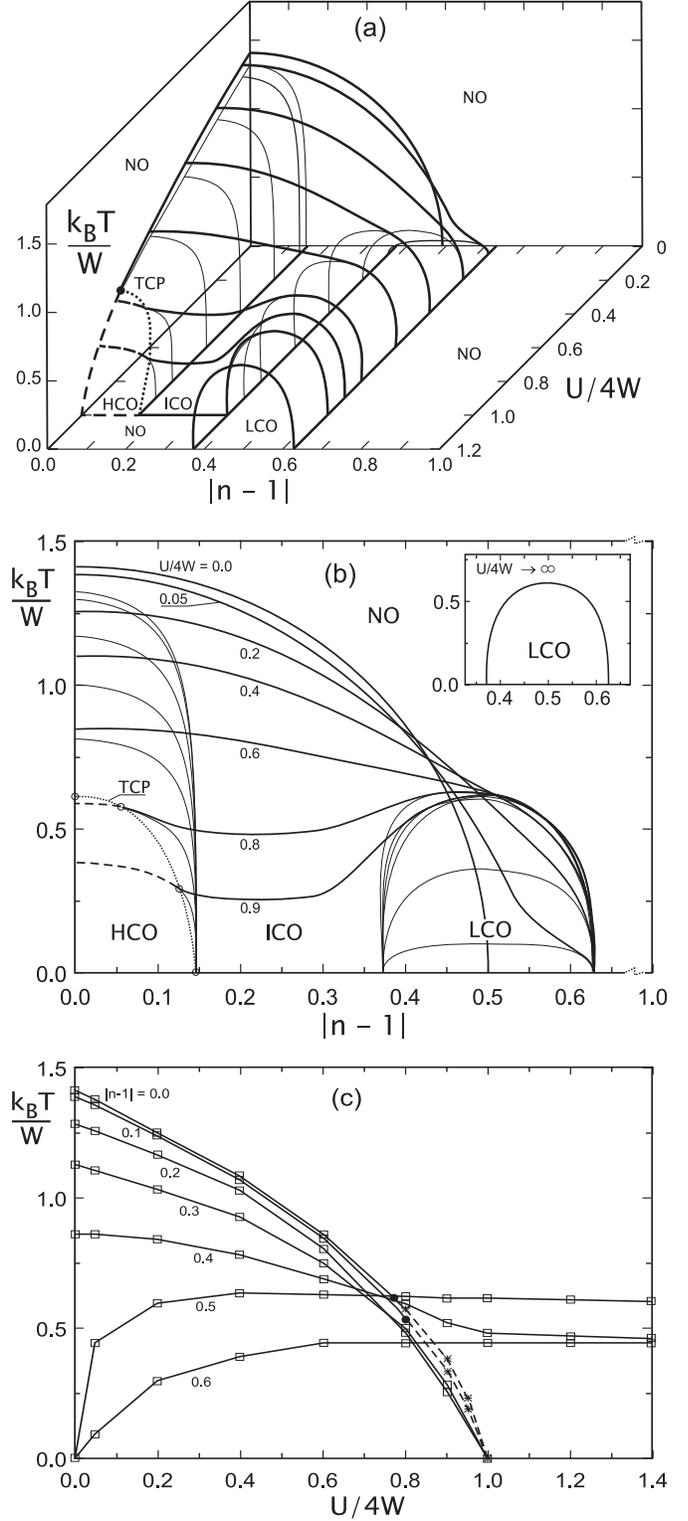}
\caption{(a) Phase diagram of model (1) as a function of $k_B T/W$, $|n-1|$ and $U/4W$. Cross-section: (b) for fixed $U/4W$ values (numbers to the curves), (c) for given $|n - 1|$ values (numbers to the curves). Bold lines indicate the 2nd-order phase transitions, whereas thin lines- the order-order boundaries. Dashed and dotted lines indicate the 1st-order and TCP lines, respectively.}
\label{3D} 
\end{figure}

Fig. \ref{3D} presents a full phase diagram in the plane $|n - 1|$ vs $U/4W$ vs $k_B T/W$. For infinite U (Fig. \ref{3D}b inset) we observe only the LCO phase. Note, that for infinite on-site repulsion only empty and singly occupied sites (for $n \leq 1$) can exist, which leads to the classical lattice gas limit (see Appendix also). The maximum value of the charge order critical temperature occurs at the quarter-filling (for electrons $n = 0.5$ or for holes $n = 1.5$). In the regime $1 < U/4W < \infty$ the shape of the phase transitions line does not change qualitatively. The LCO critical temperature very weakly depends on the value of the on-site interaction (cf. Fig. \ref{3D}a and \ref{3D}b inset).

The region of $0 \leq U/4W \leq 1$ is the most interesting for analysis (Fig. \ref{3D}). One can observe all ordered phases. In the range of $0.6 < U/4W < 1$ the LCO second order transition boundary (along with the structural LCO-ICO boundary) remains almost the same as for infinite U (Fig. \ref{3D}b,c). The LCO critical temperature ${T^{LCO}_c}$ goes down to zero with decreasing on-site repulsion $U/4W \rightarrow 0$.

The HCO phase can be observed for the electron concentrations $0.86 < n < 1.14$. The critical temperature ${T^{HCO}_c}$ grows with decreasing U. For $0.77 < U/4W \leq 1$ the first order transitions occur between HCO and NO states, while for $U/4W$ smaller than $0.77$ (the tricritical point for $n = 1$, Fig. \ref{n_1}a) the transitions have continuous character in the whole range of concentrations. On the TCP line (Fig. \ref{3D}a, dotted line) the 1st-, 2nd-order phase transitions and HCO-ICO order-order boundaries are joined (Figs. \ref{n_1}b, \ref{3D}b). 

In the $0 \leq U/4W \leq 1$ regime the ICO phase exists between the LCO and the HCO phases (Figs \ref{3D}a,b).
For $0.77 < U/4W < 1$, the ${T^{ICO}_c}$ takes a smaller value than  $max({T^{LCO}_c}, {T^{HCO}_c})$. For $U/4W < 0.77$ the maximum value of ${T^{ICO}_c}$ is also the highest value of critical temperatures for all charge ordered states.

\section{The equivalent models}
Nontrivial transformation from model (1) to spin model is based on the definition of equivalent pseudospin operators:
\begin{eqnarray}
S_{i}=n_{i}-1=n_{i\uparrow}+n_{i\downarrow}-1,
\label{eq:2}
\end{eqnarray}
where ${\{S_{i}\}}$ can take four values $\{-1,0^{'},0^{''},1\}$ (the single and double apostrophe indicate the double presence of $0$-state).

The form of Eq. \ref{UW} is now
\begin{eqnarray}
{\cal H}(S_{i})=\frac{1}{2}U{\sum_{i}S_{i}^{2}}+W\sum_{<ij>}S_{i}S_{j}
-\bar{\mu}\sum_{i}S_{i}+C,
\label{eq:3}
\end{eqnarray}
\begin{eqnarray}
C=-\bar{\mu}N-\frac{N}{2}(U+z_o W), \nonumber
\end{eqnarray}
where $\bar{\mu}=\mu-\frac{1}{2}U-z_o W$ ($z_o$ is the number of nearest-neighbors).

Thus, the EHM in the zero-bandwidth is equivalent to the pseudospin model with single-ion anisotropy $U/2$ and exchange interaction $W$ (antiferromagnetic if ${W > 0}$,  ferromagnetic if ${W < 0}$) in a effective magnetic field $\bar\mu$, with double 'zero' state degeneracy.

\begin{figure}
\includegraphics{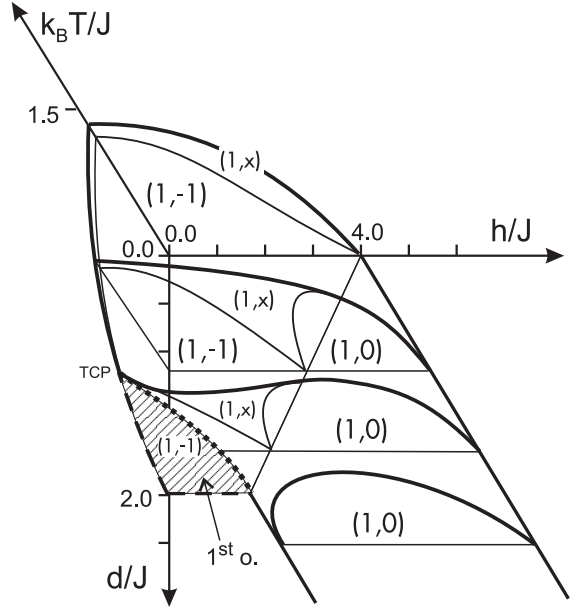}
\caption{Phase diagram of the pseudospin model (\ref{eq:3}). Notation used: $d = U/2$, $J = W$, $h = \bar{\mu}$. Thick lines indicate the 2nd-order phase transitions (antiferromagnetic $\leftrightarrow$ paramagnetic), whereas thin lines- the order-order boundaries. Dashed and dotted lines indicate the 1st-order and TCP lines, respectively.}
\label{BC3D} 
\end{figure}

Fig. \ref{BC3D} presents the phase diagrams of model (\ref{eq:3}) ($\bar{\mu} \geq 0$ corresponds to a positive value of a magnetic field in a pseudospin system). We use the canonical ensemble MC for the pseudospin (${\{S_{i}\}}$) system for a square lattice (L=60, 80) with PBC. Thick, dashed and dotted lines indicate the 2nd-, 1st-order transitions and TCP-line, respectively. Thin lines indicate structural boundaries of the pure AF phases (1,0), (1,-1) and the intermediate ordered phase (1,x) (defined as a structure of coexisting cells of pure AF phases, where $x = \{-1,0\}$). Order-order lines are determined using cluster analysis of ordered structures (analogous method as described in Sec. 2, see also our previous paper \cite{gpawlo}).

To eliminate the spin degeneracy in model (\ref{eq:3}) we map it on to a standard S=1 Ising model \cite{Micnas}. The double degeneracy of every $S_{i}=0$ leads to a factor of 2 in the partition function of the classical spin system (we def. $\{\widetilde{S}_{i}\}$ as $\{-1,0,1\}$), thus the total multiplication factor is equal:
\begin{eqnarray*}
\prod_{i}2^{1-\widetilde{S}_{i}^{2}}=2^{\sum_{i}(1-\widetilde{S}_{i}^{2})} .
\end{eqnarray*}
We rewrite the partition function of model (\ref{eq:3}) as follows:
\begin{eqnarray}
Z={\sum_{\{S_i\}=\{-1,0,0,1\}}}\exp[-\beta{\cal H}(S_{i})] =\nonumber \\
={\sum_{\{\widetilde{S}_i\}=\{-1,0,1\}}}\exp[-\beta{\cal H}(\widetilde{S}_{i})] \cdot 2^{\sum_{i}(1-\widetilde{S}_{i}^{2})} =\nonumber \\
={\sum_{\widetilde{S}_i}}\exp\lbrace-\beta [{\cal H}(\widetilde{S}_{i}) + \beta^{-1} \ln 2 \sum_{i}(\widetilde{S}_{i}^{2}-1)]\rbrace =\nonumber \\
={\sum_{\widetilde{S}_i}}\exp[-\beta{\widetilde{\cal H}}(\widetilde{S}_{i})],\nonumber \\
\end{eqnarray}
where ${\widetilde{H}}$ is the Hamiltonian of the standard S=1 Ising model, but with the effective temperature-dependent single-ion anisotropy:
\begin{eqnarray}
\widetilde{\cal H}=\left(\frac{1}{2}U+\beta^{-1}\ln 2\right){\sum_{i}\widetilde{S}_{i}^{2}} +
W\sum_{<ij>}\widetilde{S}_{i}\widetilde{S}_{j}\nonumber \\
- \bar{\mu}\sum_{i}\widetilde{S}_{i} - C',
\end{eqnarray}
where $C' = C - N\beta^{-1}\ln 2$.\\
Now, we redefine the model parameters:
\begin{eqnarray}
\Delta = \frac{1}{2}U+k_B T\ln 2, \label{delta_BC}\\
J = W, \\
H = \bar{\mu} = \mu-\frac{1}{2}U-z_o W,
\label{transf}
\end{eqnarray}
and finally we obtain the classical Blume-Capel model with $\{\widetilde{S}_i\} = \{-1,0,1\}$ in the magnetic field:
\begin{eqnarray}
{\cal H_{BC}}=\Delta{\sum_{i}\widetilde{S}_{i}^{2}}+J\sum_{<ij>}\widetilde {S}_{i}\widetilde{S}_{j}-H\sum_{i}\widetilde{S}_{i}.
\label{B-C}
\end{eqnarray}

The phase diagram of model (\ref{eq:3}) is similar to that of the well-known BC model (\ref{B-C}) (cf. phase diagram in \cite{Kimel}). From Eq. (\ref{delta_BC}) we see that the results are identical in the ground state when $\Delta = U/2$. More complicated situation occurs in finite temperatures, the critical points of model (\ref{B-C}) are shifted by about $k_B T \ln 2$ in the $T=const$ plane relative to corresponding points of model (\ref{eq:3}) (cf. Fig. \ref{n_1}a). Our results are in good agreement with \cite{Kimel} - \cite{Xavier}.

\section{Summary}

The Monte Carlo studies of the AL-EHM and the equivalent Ising S=1 AF model in the presence of magnetic field are performed for 2D square lattice with PBC. 
The grand canonical ensemble is used for the electron model, while the canonical one for the spin model.
The results of both models are complementary. 
The phase diagrams of the models analyzed include the first- and second-order transitions and tricritical points. It is shown that the ordered state contains three types of staggered phases (CO in Al-EHM and AF in the spin model) in a wide range of model parameters (e.g. the electron concentration and chemical potential).

It is undoubtedly important to discuss the physical reflection of the charge orderings in the  incommensurate fillings (beyond quarter- and half-filling): are these states stable?
The response depends on the value of the hopping $t_{ij}$, which in this paper has been arbitrarily assumed zero. The case of insignificant mobility in the ground state leads to the charge orderings in the domains and in higher temperatures to the appearance of a stable mixed phase.
For $t=0$, the CO states beyond commensurate fillings correspond to the stable uncompensated antiferromagnetism in the equivalent spin model following from the quantitative excess of spins in one direction in the presence of magnetic field.
In real systems the bandwidth is different from zero. The question is whether the analysis for a finite $t$ changes qualitatively or only quantitatively the obtained results.
The analysis for a small hopping can lead to weak magnetic correlations (the effect of the kinetic exchange $\sim 2t_{ij}^2/U$) which do not modify the present charge ordering. The quantum fluctuations exert the greatest influence on the ground state but do not change qualitatively the character of the phase diagrams at finite temperatures. 
For higher $t$ the situation drastically changes. 
The mobility of the electrons (or holes) will suppress the staggered order.
The high-temperature series expansion method \cite{Bartkowiak} has shown that for positive $U$ the presence of the hopping term acts to destabilize the charge-ordered phase and the tricritical point moves towards smaller U and higher temperatures.
Since the finite band case can only suppress the range of the existence of charge ordered phases, our new results of AL-EHM obtained in the whole range of electron concentration show the scale of charge order presence in the widest case.

In this paper we analyze the order-order boundaries for the first time. The multiphase structure of finite-temperature phase diagram is presented. The cluster analysis illustrates the microscopic mechanism of the transition from order to disorder, especially the difference between 1st- and 2nd-order phase transition. Moreover, in low temperatures the existence of the state with phase separation of different charge ordered phases is proved.

We expect that detailed results concerning the critical behavior (the critical exponents) in AL-EHM and BC models will yield no identity. For the AL-EHM we obtain results which are rather in the class of the percolation transition problem, while the equivalent spin model is in the Ising-type class.

\section{Acknowledgments}

The Author is grateful to S. Robaszkiewicz for very important suggestions and many useful discussions. The important advice by T. Kostyrko and P. Tomczak are also gratefully acknowledged.

This work was supported in part by the Polish State Committee for Scientific Research, Grant  No: 1 P03B 084 26; 2004-2006.

\section{Appendix}

For the sake of comparison with our results we present here selected solutions for AL-EHM obtained by different methods (MFA, BPA).

\begin{figure}
\includegraphics{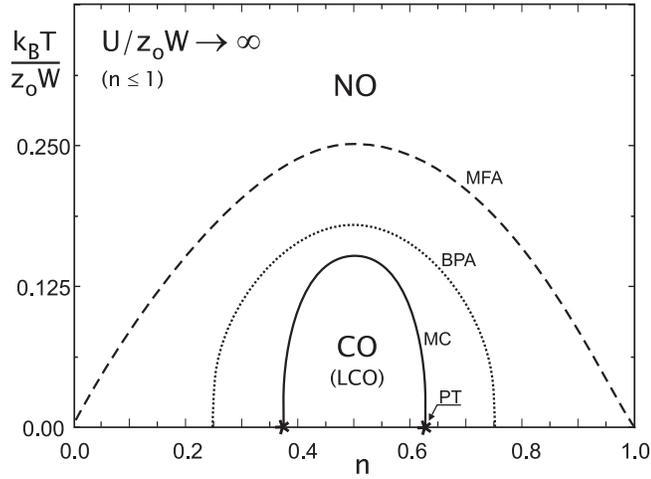}
\caption{The case of $U/z_o W \rightarrow  \infty $ ($n \leq 1$). The lines indicate critical temperatures obtained by MFA (dashed), BPA (dotted) and MC (solid). The stars indicate the percolation threshold (PT) at the ground state.}
\label{App1}
\end{figure}

In the limit of $U \rightarrow \infty$ the model is equivalent to the classical two-component lattice gas where $n_i = \{ 0, \uparrow, \downarrow \}$ (for $n \leq 1$). In the absence of external magnetic field the electron gas can be treated as the charged one-particle classical lattice gas \cite{Binder}. In this case, the MC results can be compared with the analytical results. The solutions concerning the critical temperature are for BPA \cite{Rob}
\begin{eqnarray}
k_B T_c ^{BPA} = W / \ln \frac{(q - n)^2}{(1 - 2n)q + q^2}, \nonumber
\end{eqnarray}
where $q = n(z_o n - 1) / (z_o - 1)$, and for MFA \cite{Micnas}
\begin{eqnarray}
k_B T_c ^{MFA} = z_o W n(1 - n)\nonumber
\end{eqnarray}
$n$ is the electron concentration and $z_o$ is the number of nearest-neighbors. The range of charge order changes depends on the method used (cf. Fig. \ref{App1} for $n \leq 1$). For the MFA we observe that the ordered phase exists for any electron concentration, while for BPA and MC there is a limited range of $n$ with a stable charge order (note that on the ground state the percolation threshold $\theta=0.371$ \cite{Binder} and the MC results $n=0.37$ coincide). The maximum critical temperature is always for the quarter-filling.

\begin{figure}
\includegraphics{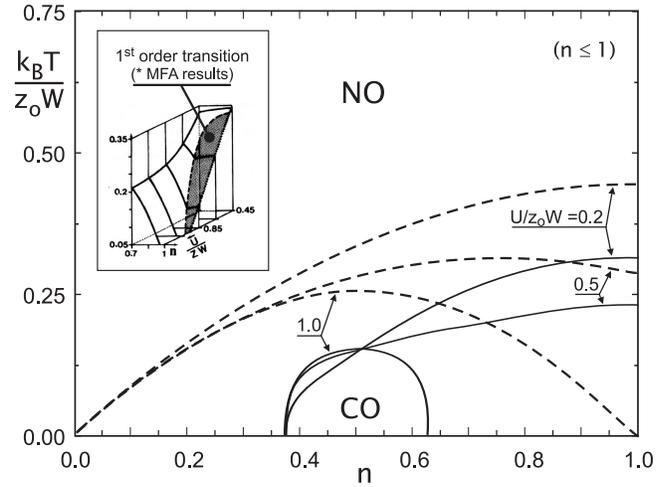}
\caption{Phase diagrams of the model (1) for $n \leq 1$ obtained using MFA (dashed lines) and MC method (solid lines)  for several $U/z_o W$. The box shows solutions for 1st-order transition (MFA, * after {\cite{Micnas}}).}
\label{App2}
\end{figure}

The most comprehensive analysis for the whole range of $U/z_o W$ has been made by R. Micnas, S. Robaszkiewicz, K. A. Chao {\cite{Micnas}} using the mean field approximation. In Fig. \ref{App2} we present the phase diagrams obtained using MFA (dashed lines) and MC (solid lines) method for several value of $U/z_o W$.
Our MC results for 2D are qualitatively comparable with MFA solutions (which are exact in $D\rightarrow \infty$). We observe analogous character of phase transition, although the range of appearance of continuous and discontinuous transitions is different (cf. Fig. 7 and 10). The 1st-order phase transitions take place at $0.5 < U/z_o W \leq 1$ for MFA analysis, while the range of $0.77 < U/4W \leq 1$ for discontinuous transition is found using MC simulation. Irrespective of the method used we observe the Mott transition for $U/ z_o W = 1$ between HCO (2020) and the NO state (1111).

\begin{thebibliography}{}

\bibitem{Gabovich} A.M. Gabovich, A.I. Voitenko, M. Ausloos, Phys. Rep. \textbf{367}, 583 (2002).
\bibitem{Goto} T. Goto, B. L{\"u}thi, Adv. in Phys. \textbf{52}, 67 (2003).
\bibitem{Reis} M. S. Reis, V. S. Amaral, J. P. Araujo, P. B. Tavares, A. M. Gomes, I. S. Oliveira, Phys. Rev. B \textbf{71}, 144413 (2005).

\bibitem{Rob_73} S. Robaszkiewicz, Phys. Stat. Sol. (b) \textbf{59}, K63 (1973); Acta. Phys. Pol. A \textbf{45}, 289 (1974).
\bibitem{Le} Duc Anh Le, Anh Tuan Hoang, Toan Thang Nguyen, Mod. Phys. Lett. B \textbf{17}, Nos. 20-21, 1103 (2003).
\bibitem{Oles} K. Ro{\'s}ciszewski, A. Ole{\'s}, J. Phys. Condens. Matt. \textbf{15}, 8363 (2003).
\bibitem{Aichhorn} M. Aichhorn, H. G. Evertz, W. von der Linden, M. Potthoff, Phys. Rev. B \textbf{70}, 235107 (2004).
\bibitem{Zang} Y. Z. Zhang, Tran Minh-Tien, V. Yushankhai, P. Thalmeier, Eur. Phys. J. B \textbf{44}, 265 (2005).
\bibitem{Hoang} A.T. Hoang, P. Thalmeier, J. Phys.: Condens. Matter \textbf{14}, 6639 (2002).
\bibitem{Tong} N-H. Tong, S-Q. Shen, R. Bulla, Phys. Rev. B \textbf{70}, 085118 (2004).
\bibitem{Onari} S. Onari, R. Arita, K. Kuroki, H. Aoki, Phys. Rev. B \textbf{70}, 094523 (2004).
\bibitem{Bari} R. A. Bari, Phys. Rev. B \textbf{3}, 2662 (1971).
\bibitem{Ihle} D. Ihle, B. Lorenz, Phys. Stat. Sol. (b) \textbf{60}, 319 (1973).
\bibitem{Beni}G. Beni, P. Pincus, Phys. Rev. B \textbf{9}, 2963 (1974).
\bibitem{Tu} R. S. Tu, T. A. Kaplan, Phys. Stat. Sol. (b) \textbf{63}, 659 (1974).
\bibitem{Lorentz} B. Lorenz, Phys. Stat. Sol. (b) \textbf{106}, K17 (1981).
\bibitem{Rice} T.M.Rice and L.Sneddon, Phys. Rev. Lett. \textbf{47}, 689 (1981).
\bibitem{Kostyrko} S. Robaszkiewicz, T. Kostyrko, Physica B \& C \textbf{112}, 389 (1982).
\bibitem{Micnas} R. Micnas, S. Robaszkiewicz, K. A. Chao, Phys. Rev. B \textbf{29}, 2784 (1984).
\bibitem{Bursil} R. J. Bursill, C. J. Thompson, J. Phys. A: Math. Gen. \textbf{26}, 4497 (1993).
\bibitem{Jedrzejewski} J. J{\c e}drzejewski, Physica A \textbf{205}, 702 (1994).
\bibitem{Bartkowiak} M. Bartkowiak, J.A. Henderson, J. Oitmaa, P.E. de Brito, Phys. Rev. B \textbf{51}, 14077 (1995).
\bibitem{Borgs} C. Borgs, J. J{\c e}drzejewski, R. Kotecky, J. Phys. A: Math. Gen. \textbf{29}, 733 (1996).
\bibitem{Halvorsen} E. Halvorsen, M. Bartkowiak, Phys. Rev B \textbf{63}, 014403 (2000).
\bibitem{Mancini} F. Mancini, Eur. Phys. J. B \textbf{47}, 527 (2005).
\bibitem{Landau} D. P. Landau, K. Binder, \textit{A Guide to Monte Carlo Simulations in Statistical Physics} (Cambridge University Press, Cambridge 2000).
\bibitem{Loison} D. Loison, C.L. Qin, K.D. Schotte, X.F Jin, Eur. Phys. J B \textbf{41}, 395 (2004).
\bibitem{Binder} K. Binder, D. P. Landau, Phys. Rev. B \textbf{21}, 1941 (1980).
\bibitem{Kimel} J.D. Kimel, S. Black, P. Carter, Yung-Li Wang, Phys. Rev. B \textbf{35}, 3347 (1987).
\bibitem{Kimel2} J.D. Kimel, Per Arne Rikvold, Yung-Li Wang, Phys. Rev. B \textbf{45}, 7237 (1992).
\bibitem{Xavier} the results for ferromagnetic BC model: J.C. Xavier, F.C. Alcaraz, D. Pena Lara, J.A. Plascak, Phys. Rev. B \textbf{57}, 11575 (1998); C.J. Silva, A.A. Caparica, J.A. Plascak, Phys. Rev. E \textbf{73}, 036702 (2006).
\bibitem{Hoshen} J. Hoshen, R. Kopelman, Phys. Rev. B \textbf{14}, 3438 (1976).
\bibitem{gpawlo} G. Paw{\l}owski, Phys. Stat. Solidi (b) \textbf{243}, 331 (2006).
\bibitem{Kirkpatrick} S. Kirkpatrick, In \textit{III-Condensed Matter}, ed. R. Balian, R. Maynard and G. Toulouse (North-Holland, Amsterdam 1979) p.321.
\bibitem{Rob} S. Robaszkiewicz, private communication. See also: Acta. Phys. Pol. A \textbf{85}, 117 (1993).
\end{thebibliography}
\end{document}